\documentclass[12pt]{article}
\usepackage{fullpage}
\usepackage{cite}
\usepackage{graphicx}
\usepackage{amsmath}
\usepackage{amssymb}
\usepackage{bm}
\title{}
\date{}
\renewcommand{\vec}[1]{\boldsymbol{#1}}

\begin{document}
\bibliographystyle{utphys}
\newcommand{\msbar}{\ensuremath{\overline{\text{MS}}}}
\newcommand{\DIS}{\ensuremath{\text{DIS}}}
\newcommand{\abar}{\ensuremath{\bar{\alpha}_S}}
\newcommand{\bb}{\ensuremath{\bar{\beta}_0}}
\newcommand{\rc}{\ensuremath{r_{\text{cut}}}}
\newcommand{\Nd}{\ensuremath{N_{\text{d.o.f.}}}}
\setlength{\parindent}{0pt}

\titlepage
\begin{flushright}
CERN-TH-2016-027, Edinburgh 2016/02 \\
\end{flushright}

\vspace*{0.5cm}

\begin{center}
{\Large \bf The double copy: Bremsstrahlung and accelerating black holes}
\end{center}

\vspace*{1cm}

\begin{center} 
\textsc{Andr\'{e}s Luna$^a$\footnote{a.luna-godoy.1@research.gla.ac.uk}, 
Ricardo Monteiro$^b$\footnote{ricardo.monteiro@cern.ch}, 
Isobel Nicholson$^c$\footnote{i.nicholson@sms.ed.ac.uk}, \\
Donal O'Connell$^{c,d}$\footnote{donal@staffmail.ed.ac.uk},
 and 
Chris D. White$^a$\footnote{Christopher.White@glasgow.ac.uk} } \\

\vspace*{0.5cm} $^a$ School of Physics and Astronomy, University of Glasgow,\\ Glasgow G12 8QQ, Scotland, UK\\

\vspace*{0.5cm} $^b$ Theoretical Physics Department, CERN, Geneva, Switzerland\\

\vspace*{0.5cm} $^c$ Higgs Centre for Theoretical Physics, \\School of Physics and Astronomy, The University of Edinburgh,\\
Edinburgh EH9 3JZ, Scotland, UK\\

\vspace*{0.5cm} $^d$ Kavli Institute for Theoretical Physics,
University of California,\\
Santa Barbara, CA 93106-4030 USA\\

\end{center}

\vspace*{0.5cm} 

\begin{abstract}
\noindent
Advances in our understanding of perturbation theory suggest the existence of a correspondence
between classical general relativity and Yang-Mills theory. A
concrete example of this correspondence, which is known as the double copy, was recently introduced for the case of stationary Kerr-Schild spacetimes.
Building on this foundation, we
examine the simple time-dependent case of an accelerating, radiating point source.
The gravitational
solution, which generalises the Schwarzschild solution, includes a non-trivial stress-energy tensor. This stress-energy tensor corresponds to a gauge theoretic current in the double copy. We
interpret both of these sources as representing the radiative part of
the field. Furthermore, in the simple example of Bremsstrahlung, we
determine a scattering amplitude describing the radiation, maintaining the double copy throughout.
Our results
provide the strongest evidence yet that the classical double copy is
directly related to the BCJ double copy for scattering
amplitudes.
\end{abstract}
\newpage

\section{Introduction}
\label{sec:intro}

Our most refined understanding of nature is founded on two major theoretical frameworks: general relativity and Yang-Mills theory. There is much in common between these two: local symmetries play an important role in their structure; there are simple action principles for both theories; the geometry of fibre bundles is common to the physical interpretation of the theories. But at the perturbative level, general relativity seems to be a vastly different creature to Yang-Mills theory. Indeed, the Einstein-Hilbert Lagrangian, when expanded in deviations of the spacetime metric from some fiducial metric (such as the Minkowski metric) contains terms with arbitrarily many powers of the deviations. This is in stark contrast to the Yang-Mills Lagrangian, which contains at most fourth order terms in perturbation theory. \\

From this perturbative point of view, it is therefore remarkable that Kawai, Lewellen and Tye (KLT) found~\cite{Kawai:1985xq} that every tree scattering amplitude in general relativity can be expressed as a sum over products of two colour-stripped Yang-Mills scattering amplitudes. 
Therefore the KLT relations and the Yang-Mills Lagrangian together can be used to reconstruct the Lagrangian of general relativity~\cite{Bern:1999ji}. This suggests that there may be a KLT-like map between solutions of general relativity and solutions of Yang-Mills theory. \\

More recently, the perturbative relationship between gauge and gravity theories has been formulated in a particularly suggestive manner by Bern, Carrasco and Johansson (BCJ)~\cite{Bern:2008qj,Bern:2010ue,Bern:2010yg}.
BCJ found that gravity $n$-point amplitudes can be obtained from $n$-point gauge
theory counterparts at the level of diagrams. Specifically, the BCJ prescription is simply to replace the colour factor of each diagram by an additional copy of the diagram's kinematic numerator. This replacement must be performed in a particular representation of the amplitude, where the kinematic numerators satisfy the algebraic properties of the corresponding colour factor. In particular, the kinematic factors must satisfy the same Jacobi identities and antisymmetry properties as the colour factors. For this reason, the BCJ representation of the kinematic numerators is known as a colour-dual representation. The procedure of replacing colour factors in gauge theory scattering amplitudes with another copy of the kinematic numerator is known as the double copy, since it represents gravity scattering amplitudes as two copies of Yang-Mills scattering amplitudes. \\

The validity of the BCJ double copy and the existence of colour-dual numerators has been
proven at
tree-level~\cite{BjerrumBohr:2009rd,Stieberger:2009hq,Bern:2010yg,BjerrumBohr:2010zs,Feng:2010my,Tye:2010dd,Mafra:2011kj,Monteiro:2011pc,BjerrumBohr:2012mg}
(where it is equivalent to the KLT
relations~\cite{Kawai:1985xq}). One very exciting feature of the BCJ procedure is that it admits a simple extension to loop diagrams in the quantum theory~\cite{Bern:2010ue}. This extension remains conjectural, but it has been verified in highly nontrivial
examples at multiloop
level~\cite{Bern:2010ue,Bern:1998ug,Green:1982sw,Bern:1997nh,Oxburgh:2012zr,Carrasco:2011mn,Carrasco:2012ca,Mafra:2012kh,Boels:2013bi,Bjerrum-Bohr:2013iza,Bern:2013yya,Bern:2013qca,Nohle:2013bfa,Bern:2013uka,Naculich:2013xa,Du:2014uua,Mafra:2014gja,Bern:2014sna,Mafra:2015mja,He:2015wgf,Bern:2015ooa,Mogull:2015adi,Chiodaroli:2015rdg}. All-order
evidence can be obtained in special kinematic
regimes~\cite{Oxburgh:2012zr,Saotome:2012vy,Vera:2012ds,Johansson:2013nsa,Johansson:2013aca},
but a full proof of the correspondence has to date been missing (see,
however,
refs.~\cite{Monteiro:2013rya,Tolotti:2013caa,Fu:2013qna,Du:2013sha,Fu:2012uy,Naculich:2014naa,Naculich:2014rta,Chiodaroli:2014xia,Carrasco:2013ypa,Litsey:2013jfa,Nagy:2014jza,Anastasiou:2015vba,Johansson:2015oia,Lee:2015upy,Barnett:2014era}
for related studies).\\

Motivated by this progress, a double copy for classical field solutions (which we will refer to as the classical double copy) has been proposed~\cite{Monteiro:2014cda}. This classical double copy is similar in structure to the BCJ double copy for scattering amplitudes: in both cases, the tensor structure of gravity is constructed from two copies of the vector structure of gauge theory. In addition, scalar propagators are present in both cases; these scalars are exactly the same in gauge and gravitational processes.
However, the classical double copy~\cite{Monteiro:2014cda} is only understood at present for
the special class of Kerr-Schild solutions in general
relativity. This reflects the particularly simple structure of Kerr-Schild metrics: the Kerr-Schild ansatz has the remarkable property that the Einstein equations exactly linearise. Therefore we can anticipate that any Yang-Mills solution related to a Kerr-Schild spacetime must be particularly simple. Indeed, the authors of~\cite{Monteiro:2014cda} showed that any stationary Kerr-Schild solution has a well-defined
single copy that satisfies the Yang-Mills
equations, which also take the linearised form. 
While the structure of the classical double copy is very reminiscent of the BCJ double copy, so far no precise link has been made between the two. One aim of the present article is to provide such a link.\\

Although the classical double copy is only understood for a restricted class of solutions, many of these are familiar. For example, the Schwarzschild and Kerr black holes are members of this class; in higher dimensions, the Myers-Perry black holes are included~\cite{Monteiro:2014cda}. The relationship between classical solutions holds for all stationary Kerr-Schild solutions, but other Kerr-Schild solutions are known to have appropriate single copies. A particularly striking example is the shockwave in gravity and gauge theory; the double copy of this pair of solutions was pointed out by Saotome and Akhoury~\cite{Saotome:2012vy}. In further work, the classical double copy has been extended~\cite{Luna:2015paa}
to the Taub-NUT solution~\cite{Taub,NUT}, which has a double
Kerr-Schild form and whose single copy is a dyon in gauge theory. \\

Despite this success, Kerr-Schild solutions are very
special and do not easily describe physical systems which seem very natural from the point of view of the double copy for scattering amplitudes.
For example, there is no two-form field or dilaton on the gravity side; there are no non-abelian features on the gauge theory side; the status of the sources must be better understood. In cases where the sources are point particle-like, the classical double copy relates the gauge theory current density to the
gravity energy-momentum tensor in a natural way~\cite{Monteiro:2014cda,Luna:2015paa}. For extended sources, extra pressure terms on the gravity side are needed to stabilise the matter distribution. Furthermore, reference~\cite{Ridgway:2015fdl} pointed out that in certain gravity solutions the energy-momentum tensor does not satisfy the weak and/or strong energy conditions of general relativity. \\

The aim of this paper is to extend the classical double copy of
refs.~\cite{Monteiro:2014cda,Luna:2015paa} by
considering one of the simplest situations involving explicit time
dependence, namely that of an arbitrarily accelerating, radiating
point source. We will see that this situation can indeed be interpreted in the
Kerr-Schild language, subject to the introduction of additional source
terms for which we provide a clear interpretation. One important fact
which will emerge is that these sources themselves have a double copy structure. 
We will demonstrate that the sources can
be related directly to scattering amplitudes, maintaining the double copy throughout. This provides a
direct link between the classical double copy and the BCJ procedure for
amplitudes, 
strongly bolstering the argument that these double copies are the same. The gravitational solution of interest to us is a time-dependent
generalisation of the Schwarzschild solution; we will see that this gravitational
system is a precise double copy of an accelerating point particle. Since there is a double copy
of the sources, and these describe the radiation fields, we
learn that the gravitational radiation emitted by a black hole which undergoes a short
period of acceleration is a precise double copy of electromagnetic Bremsstrahlung. \\

The structure of our paper is as follows. In
section~\ref{sec:KSreview}, we briefly review the Kerr-Schild double
copy. In section~\ref{sec:KSradiate}, we present a known
Kerr-Schild solution for an accelerating particle, before examining
its single copy. We will find that additional source terms appear in
the gauge and gravity field equations, and in section~\ref{sec:sources}
we relate these to scattering amplitudes describing radiation, by
considering the example of Bremsstrahlung. In
section~\ref{sec:energy}, we examine the well-known energy conditions
of GR for the solutions under study. Finally, we discuss our results and
conclude in section~\ref{sec:conclude}. Technical details are
contained in an appendix.

\section{Review of the Kerr-Schild double copy}
\label{sec:KSreview}

Let us begin with a brief review of the Kerr-Schild
double copy, originally proposed in~\cite{Monteiro:2014cda,Luna:2015paa}. We define the graviton field via
\begin{equation}
g_{\mu\nu}=\bar{g}_{\mu\nu}+\kappa h_{\mu\nu},\quad \kappa=\sqrt{16\pi G_\textrm{\tiny{N}}}
\label{gravdef}
\end{equation}
where $G_\textrm{\tiny{N}}$ is Newton's constant, and $\bar{g}_{\mu\nu}$ is a
background metric, which, for the purposes of the present paper, we will take to be the Minkowski metric.\footnote{We choose to work with a negative signature metric $\eta = \textrm{diag}(1,-1,-1,-1)$.} There is a special class of {\it Kerr-Schild}
solutions of the Einstein equations, in which the graviton
has the form
\begin{equation}
h_{\mu\nu}= - \frac\kappa 2 \phi k_\mu k_\nu,
\label{KSdef}
\end{equation}
consisting of a scalar function $\phi$ multiplying the outer product
of a vector $k_\mu$ with itself. We have inserted a negative sign in this definition for later convenience. The vector $k_\mu$ must be null and
geodesic with respect to the background:
\begin{equation}
\bar{g}_{\mu\nu} \,k^\mu\, k^\nu=0,\quad  (k\cdot D) k=0,
\label{geodetic}
\end{equation}
where $D^\mu$ is the covariant derivative with respect to the
background metric. It follows that $k_\mu$ is also null and geodesic with respect to the metric $g_{\mu\nu}$. These solutions have the remarkable property that
the Ricci tensor with mixed upstairs / downstairs indices is {\it
  linear} in the graviton. More specifically, one has 
\begin{equation}
R^\mu_{\ \nu}=\bar{R}^\mu_{\ \nu}-\kappa \left[h^{\mu}_{\ \rho}\bar{R}^\rho_{\ \nu}
-\frac{1}{2} D_\rho\left(D_\nu h^{\mu\rho}+D^\mu h^\rho_{\ \nu}
-D^\rho h^\mu_{\ \nu}\right)\right],
\label{Ricci}
\end{equation}
where $\bar{R}_{\mu\nu}$ is the Ricci tensor associated with
$\bar{g}_{\mu\nu}$, and we have used the fact that $h^\mu_{\ \mu} = 0$. It follows that the Einstein equations themselves
linearise. Furthermore, ref.~\cite{Monteiro:2014cda} showed that for
every stationary Kerr-Schild solution (i.e. where neither $\phi$ nor
$k^\mu$ has explicit time dependence), the gauge field
\begin{equation}
A^a_\mu=c^a \phi\,k^\mu,
\label{AmuKS}
\end{equation}
for a constant colour vector $c^a$, solves the Yang-Mills
equations. Analogously to the gravitational case, these equations take a linearised form due to the trivial colour dependence of the solution. We then refer to such a gauge field as the {\it single
  copy} of the graviton $h_{\mu\nu}$, since it involves only one
factor of the Kerr-Schild vector $k_\mu$ rather than two. Note that
the scalar field $\phi$ is left untouched by this procedure. This was
motivated in ref.~\cite{Monteiro:2014cda} by taking the {\it zeroth
  copy} of eq.~(\ref{AmuKS}) (i.e. stripping off the remaining $k^\mu$
factor), which leaves the scalar field itself. The zeroth copy of a
Yang-Mills theory is a biadjoint scalar field theory, and the field
equation linearises for the scalar field obtained from
eq.~(\ref{AmuKS}). The scalar function $\phi$ then corresponds to a
propagator, and is analogous to the untouched denominators (themselves
scalar propagators) in the BCJ double copy for scattering
amplitudes.\\

Source terms for the biadjoint, gauge and gravity theories also match
up in a natural way in the Kerr-Schild double copy. Pointlike sources
in a gauge theory map to point particles in gravity, where electric
and (monopole) magnetic charge are replaced by mass and NUT charge
respectively~\cite{Luna:2015paa}. Extended source distributions (such
as that for the Kerr black hole considered in
ref.~\cite{Monteiro:2014cda}) lead to additional pressure terms in the
gravity theory, which are needed to stabilise the source distribution
so as to be consistent with a stationary solution. Conceptual
questions relating to extended source distributions have been further
considered in ref.~\cite{Ridgway:2015fdl}, regarding the well-known energy conditions of general
relativity. In this work, we will consider point-like objects throughout, and 
therefore issues relating to extended source distributions will not trouble us.
Nevertheless we will discuss the energy conditions in section~\ref{sec:energy} below. \\

Let us emphasise that the Kerr-Schild double
copy cannot be the most general relationship between solutions in gauge
and gravity theories. Indeed, the field one obtains upon taking the
outer product of $k^\mu$ with itself is manifestly
symmetric. Moreover, the null condition on $k^\mu$ means that the
trace of the field vanishes. Hence, the Kerr-Schild double copy is
unable to describe situations in which a two-form and / or dilaton are
active in the gravity theory. This contrasts sharply with the double copy procedure for scattering amplitudes, which easily incorporates these fields.
Furthermore, Yang-Mills amplitudes only
obey the double copy when written in BCJ dual form, meaning that
certain Jacobi relations are satisfied by the kinematic numerator
functions~\cite{Bern:2008qj,Bern:2010ue,Bern:2010yg}. It is not known
what the analogue of this property is in the classical double copy
procedure. All of these considerations suggest that the Kerr-Schild
story forms part of a larger picture, and in order to explore this it
is instructive to seek well-defined generalisations of the results of
refs.~\cite{Monteiro:2014cda,Luna:2015paa}.

\section{Kerr-Schild description of an accelerating point particle}
\label{sec:KSradiate}

In this article, we will go beyond previous work on the Kerr-Schild double copy~\cite{Monteiro:2014cda,Luna:2015paa} by considering an accelerating point particle. This is a particularly attractive case, because an accelerating point particle must radiate, so we may hope to make direct contact between the double copy for scattering amplitudes and for Kerr-Schild backgrounds. We first describe a well-known Kerr-Schild spacetime containing an accelerating point particle, before constructing the associated single-copy gauge theoretic solution. We find that the physics of the single copy is particularly clear, allowing a refined understanding of the gravitational system. We will build on this understanding in section~\ref{sec:sources} to construct a double copy pair of scattering amplitudes from our pair of Kerr-Schild solutions in gauge theory and gravity in a manner that preserves the double copy throughout.

\subsection{Gravity solution}

Consider a particle of mass $M$ following an arbitrary timelike worldline $y(\tau)$, parameterised by its proper time $\tau$ so that the proper velocity of the particle is the tangent to the curve
\begin{equation}
\lambda^\mu=\frac{d y^\mu}{d\tau}.
\label{lambdadef}
\end{equation}
An exact Kerr-Schild spacetime containing this massive accelerating particle is known,
though the spacetime contains
an additional stress-energy tensor; we will understand the physical role of this stress-energy tensor below.
\begin{figure}
\begin{center}
\scalebox{0.7}{\includegraphics{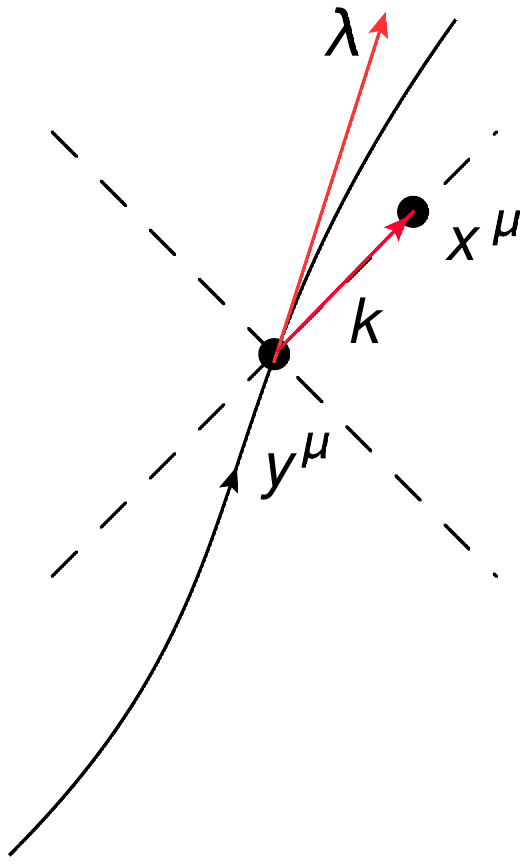}}
\caption{Geometric interpretation of the Kerr-Schild solution for an accelerated particle.}
\label{fig:worldline}
\end{center}
\end{figure}
A useful geometric interpretation of the null vector $k_\mu$ appearing in the solution has
been given in refs.~\cite{Newman-Unti,Kinnersley:1970zw,Vaidya:1972zz}
(see ref.~\cite{Stephani:2003tm} for a review), as follows. Given an
arbitrary point $y^\mu(\tau)$ on the particle worldline, one may draw a
light cone as shown in figure~\ref{fig:worldline}.  At all points
$x^\mu$ along the light-cone, one may then define the null vector
\begin{equation}
k^\mu(x)=\left.\frac{(x-y(\tau))^\mu}{r}\right|_\mathrm{ret},\quad r=\left.\lambda\cdot (x-y)\right|_\mathrm{ret},
\label{kdef}
\end{equation}
where the instruction ret indicates that $y$ and $\lambda$ should be evaluated at the retarded time $\tau_\mathrm{ret}$, i.e. the
value of $\tau$ at which a past light cone from $x^\mu$ intersects the
worldline. 
Calculations are facilitated by noting that:
\begin{align}
\label{useful1}
\partial_\mu k_\nu &= \partial_\nu k_\mu = \frac{1}{r} \left( \eta_{\mu\nu} -\lambda_\mu k_\nu -k_\mu \lambda_\nu - k_\mu k_\nu\, (-1+r k\cdot \dot{\lambda}) \right) , \\
\partial_\mu r &= \lambda_\mu + k_\mu (-1+r k\cdot \dot{\lambda}),
\label{useful2}
\end{align}
where dots denote differentiation with respect to the proper time $\tau$. \\

The Kerr-Schild metric associated with this particle is
\begin{equation}
g_{\mu\nu} = \eta_{\mu\nu} - \frac{\kappa^2}{2} \phi k_\mu k_\nu
\end{equation}
where $k_\mu$ is precisely the vector of eq.~\eqref{kdef}
and different functional forms for $\phi$ lead to different solutions. The
scalar function corresponding to an accelerating particle is given by
\begin{equation}
\phi=\frac{M}{4\pi r}.
\label{phidef}
\end{equation}
Plugging this into the Einstein equations, one
finds
\begin{equation}
{G^\mu}_\nu\equiv {R^\mu}_\nu-\frac{R}{2}{\delta^\mu}_\nu=\frac{\kappa^2}{2}
{{T_\textrm{\tiny{KS}}}^\mu}_\nu,
\label{GmunuKS}
\end{equation}
where\footnote{We note what appears to be a typographical error in
  ref.~\cite{Stephani:2003tm}, where the energy-momentum tensor
  contains an overall factor of 4 rather than 3. We have explicitly
  carried out the calculation leading to eq.~(\ref{GmunuKS}), and
  found agreement with
  refs.~\cite{Newman-Unti,Kinnersley:1970zw,Vaidya:1972zz}.}
\begin{equation}
T_\textrm{\tiny{KS}}^{\mu\nu}=\left.\frac{3M}{4\pi}\frac{k\cdot \dot{\lambda}}{r^2}
k^\mu k^\nu\right|_{\rm ret}.
\label{Tmunurad}
\end{equation}
Thus, the use of Kerr-Schild coordinates for the
accelerating particle leads to the presence of a non-trivial
energy-momentum tensor on the right-hand side of the Einstein equations. We can already see that this extra term vanishes in
the stationary case ($\dot{\lambda}^\mu=0$), consistent with the
results of ref.~\cite{Monteiro:2014cda}. More generally, this stress-energy
tensor $T_\textrm{\tiny{KS}}^{\mu\nu}$ describes a pure radiation field present
in the spacetime. The physical interpretation of this source is particularly
clear in the electromagnetic ``single copy'' of this system, to which we now turn.

\subsection{Single copy}

Having examined a point particle in arbitrary motion in a Kerr-Schild spacetime, we may
apply the classical single copy of eq.~(\ref{AmuKS}) to construct a corresponding gauge 
theoretic solution. This procedure is not guaranteed to work, given that the single copy of
refs.~\cite{Monteiro:2014cda,Luna:2015paa} was only shown to apply in the case of stationary
fields. However, we will see that we can indeed make sense of the
single copy in the present context. Indeed, the physical interpretation of the stress-energy
tensor $T_\textrm{\tiny{KS}}^{\mu\nu}$ we encountered in the gravitational situation is
illuminated by the single copy. \\

The essence of the Kerr-Schild double-copy is a relationship between
gauge theoretic solutions $A^\mu = k^\mu \phi$ and Kerr-Schild
metrics which is simply expressed as $k_\mu \rightarrow k_\mu k_\nu$.
Thus, the single-copy of 
\begin{equation}
h^{\mu \nu} = -\frac {M\kappa}{2} \frac{1}{4 \pi r} k^\mu k^\nu
\end{equation}
is\footnote{In principle, one should include an arbitrary
  colour index on the field strength and current density. Given that
  the field equations are abelian, however, we ignore this. The
  resulting solution can be easily embedded in a non-abelian theory,
  as in refs.~\cite{Monteiro:2014cda,Luna:2015paa}.
  Note that the abelian character of this theory also implies that we make the replacement
$\frac{M\kappa}{2}\rightarrow g$ (cf. eq. (38) from ref. \cite{Monteiro:2014cda}).}
\begin{equation}
A^\mu = g \frac{1}{4 \pi r} k^\mu,
\label{eq:pointAmuKS}
\end{equation}
where $g$ is the coupling constant.\footnote{The relative sign between $h_{\mu\nu}$ and $A_\mu$ is necessary
in our conventions to ensure that positive masses yield attractive gravitational fields while positive scalar potentials $A^0$ are sources for electric field lines $\vec E = - \vec{\nabla} A^0$.}
Inserting this gauge field into the Yang-Mills equations, one finds that nonlinear terms vanish,
leaving the Maxwell equations
\begin{equation}
\partial^\mu F_{\mu\nu}=j_{\textrm{\tiny{KS}}\,\nu},
\label{Maxwell}
\end{equation}
where
\begin{equation}
F_{\mu\nu}=\partial_\mu A_\nu-\partial_\nu A_\mu
\label{Fmunu}
\end{equation}
is the usual electromagnetic field strength tensor. \\

A key result is that we find that the 
current density appearing in the Maxwell equations is given by
\begin{equation}
j_{\textrm{\tiny{KS}}\,\nu}=\left.2\frac{g}{4\pi}
\frac{k\cdot \dot{\lambda}}{r^2}k_\nu
\right|_\mathrm{ret}.
\label{jadef}
\end{equation}
It is important to note
that the current density $j_{\textrm{\tiny{KS}}}$ is related to the energy-momentum
tensor, eq.~(\ref{Tmunurad}), we encountered in the gravitational case. Indeed the relationship between these sources is in accordance with the Kerr-Schild double copy: it involves
a single factor of the Kerr-Schild vector $k^\mu$, with similar
prefactors, up to numerical constants. We will return to this interesting fact in the
following section. \\

The role of the Kerr-Schild current density $j_{\textrm{\tiny{KS}}}$ can be understood by examining
our single-copy gauge field, eq.~\eqref{eq:pointAmuKS}, in more detail. Let us compute
the electromagnetic field strength tensor of this system. 
Using the results \eqref{useful1} and \eqref{useful2}, it is easy to check that
\begin{equation}
F_{\mu\nu} =
\partial_\mu A_{\nu}-\partial_\nu A_{\mu}
=\frac{g}{4\pi r^2}\left(k_\mu \lambda_\nu- \lambda_\mu k_\nu\right).
\end{equation}
A first observation about this field strength tensor is that it falls off as $1/r^2$ and does not depend on the acceleration
of the particle. Therefore, it does not describe the radiation field of the accelerated point particle in empty space, since the
radiation fields must fall off as $1/r$ and are linear in the acceleration. Secondly,
this tensor is manifestly constructed from Lorentz covariant quantities. In the instantaneous rest frame of the particle, 
$\lambda^\mu = (1,0,0,0)$ and $k^\mu = (1, \hat{\vec r})$, and in this frame it is easy to see that the field strength is simply the Coulomb
field of the point charge. Therefore, in a general inertial frame, our field strength tensor describes precisely the boosted Coulomb
field of a point charge, omitting the radiation field completely. \\

The absence of radiation in the electromagnetic field strength makes the interpretation of the current density $j_{\textrm{\tiny{KS}}}$ in the Maxwell
equation obvious. This source must describe the radiation field of the point particle.
To see this more concretely, let us compare our Kerr-Schild gauge field to the standard Li\'enard-Wiechert solution $A_\textrm{\tiny{LW}}^\mu = \frac{g}{4\pi r} \lambda^\mu$,  which describes a point particle moving in an arbitrary manner in empty space (see e.g.~\cite{jackson_classical_1999}). This comparison is facilitated by defining a ``radiative gauge field''
\begin{equation}
A^{\mu}_{\rm rad}=\frac{g}{4\pi r}(\lambda^\mu-k^\mu),
\label{Amuraddef}
\end{equation}
which satisfies
\begin{equation}
F^{\mu\nu}_{\rm rad}\equiv
\partial^\mu A^{\nu}_{\rm rad}-\partial^\nu A^{\mu}_{\rm rad}
=\frac{g}{4\pi r}\left(k^\mu\beta^\nu-\beta^\mu k^\nu\right),
\label{Fmunuraddef}
\end{equation}
where
$\beta_\mu=\dot{\lambda}_\mu-\lambda_\mu k\cdot \dot{\lambda}.$
Thus, $F^{\mu\nu}_{\rm rad}$ is the radiative field strength of the point particle: it is linear in the particle acceleration, and falls off as $1/r$ at large distances. \\

Now, since the Li\'enard-Wiechert field is a solution of the vacuum Maxwell equation, we know that $\partial_\mu \left(F^{\mu\nu}+F^{\mu\nu}_{\rm rad}\right)=0$ and, consequently,
\begin{equation}
\partial_\mu F^{\mu\nu}_{\rm rad}=-j^{\nu}_\textrm{\tiny{KS}}.
\label{Maxwell2}
\end{equation}
We interpret $j_\textrm{\tiny{KS}}$ as a divergence of the radiative
field strength: we have put the radiation
part of the gauge field on the right-hand side on the Maxwell equations, rather
than the left.  \\

Let us now summarise what has happened. By choosing Kerr-Schild
coordinates for the accelerating particle in gravity, an extra
energy-momentum tensor $T_\textrm{\tiny{KS}}^{\mu\nu}$ appeared on the right-hand side of the Einstein
equations. The single copy turns an energy density into a charge
density (as in
refs.~\cite{Monteiro:2014cda,Luna:2015paa,Ridgway:2015fdl}). Thus, the
energy-momentum tensor in the gravity theory becomes a charge current $j^\mu_\textrm{\tiny{KS}}$
in the gauge theory. We have now seen that this current represents the
radiation coming from the accelerating charged particle, and this also allows
us to interpret the corresponding energy-momentum tensor on the gravity side: it
represents gravitational radiation from an accelerating point mass. \\

Indeed, our use of Kerr-Schild coordinates forced the radiation to appear
in this form. The vector $k_\mu$ which is so crucial for our approach
is twist-free: $\partial_\mu k_\nu = \partial_\nu k_\mu$. It is known that
twist-free, vacuum, Kerr-Schild metrics are of Petrov type D, and therefore
there is no gravitational radiation in the metric; see ref.~\cite{Stephani:2003tm} for a review. Correspondingly, the radiation
is described by the Kerr-Schild sources. \\

The radiation fields of the accelerating charge in gauge theory, and the accelerating point mass in gravity, are described
in Kerr-Schild coordinates by sources $j^\mu_\textrm{\tiny{KS}}$ and $T^{\mu\nu}_\textrm{\tiny{KS}}$. The structure of
these sources reflects the Kerr-Schild double copy procedure: up to numerical factors, one replaces the vector $k_\mu$ 
by the symmetric trace-free tensor $k_\mu k_\nu$ to pass from gauge theory to gravity. This relationship between the
sources, which describe radiation, is highly suggestive. Indeed, it is a standard fact that scattering amplitudes can be obtained
from (amputated) currents. We may therefore anticipate that the structural relationship between the Kerr-Schild currents is
related to the standard double copy for scattering amplitudes. \\

Nevertheless, there are still some puzzles regarding the 
analysis above. What, for example, are we to make of the different numerical
factors appearing in the definitions eqs.~(\ref{Tmunurad}) and~(\ref{jadef}) of
the Kerr-Schild stress tensor and current density? If these sources are related
to amplitudes, we expect a double copy which is local in momentum space. How
can our currents be local in position space?
More generally, how can we be sure that the Kerr-Schild double copy is
indeed related to the standard BCJ procedure? The
answer to these questions is addressed in the following
section, in which we interpret the radiative sources directly in terms of
scattering amplitudes. \\

Before proceeding, however, let us comment on the physical interpretation of the particle in the solutions under study. We considered how the particle affects the gauge or gravity fields, but we did not consider the cause of the acceleration of the particle, i.e.~its own equation of motion. In the standard Li\'enard-Wiechert solution, the acceleration is due to a background field. It is therefore required that this background field does not interact with the radiation, otherwise the solution is not valid. This is true in electromagnetism or in its embedding in Yang-Mills theory. However, in the gravity case, one cannot envisage such a situation. Therefore, one should think of this particle merely as a boundary condition, and not as a physical particle subject to forces which would inevitably affect the Einstein equations. What we are describing here is a mathematical map between solutions in gauge theory and gravity, a map which exists irrespective of physical requirements on the solutions. In a similar vein, ref.~\cite{Ridgway:2015fdl} showed that energy-momentum tensors obtained through the classical double copy do not necessarily obey the positivity of energy conditions in general relativity.

\section{From Kerr-Schild sources to amplitudes}
\label{sec:sources}

In the previous section, we saw that the Kerr-Schild double copy can
indeed describe radiating particles. The radiation appears as a
source term on the right-hand side of the field equations. In this section, we
consider a special case of this radiation, namely Bremsstrahlung
associated with a sudden rapid change in direction. 
By Fourier transforming the source terms in the gauge and gravity
theory to momentum space, we will see that they directly yield known scattering
amplitudes which manifestly double copy. Moreover, the manipulations required to extract the scattering amplitudes
in gauge theory and in gravity are precisely parallel. We will preserve the double copy structure at each step, so that
the double copy property of the scattering amplitudes emerges from the $k_\mu \rightarrow k_\mu k_\nu$ structure of the Kerr-Schild double copy.
In this way, we firmly establish a link between the
classical double copy and the BCJ double copy of scattering amplitudes.\\

In order to study Bremsstrahlung, we consider a particle 
which moves with velocity
\begin{equation}
\lambda^\mu(\tau) = u^\mu + f(\tau)({u'}^\mu-u^\mu),
\end{equation}
where
\begin{equation}
f(\tau)=\left\{\begin{array}{ll}0,&\quad \tau< -\epsilon\\ 
1,&\quad \tau> \epsilon \end{array}   \right.
\label{frange}
\end{equation}
and, in the interval $(-\epsilon,\epsilon)$, $f(\tau)$ is smooth but otherwise arbitrary. This describes a particle which moves with constant
velocity $\lambda^\mu=u^\mu$ for $\tau<-\epsilon$, while for $\tau>\epsilon$ the particle moves with a different constant velocity $\lambda^\mu = u'^\mu$.
Thus, the particle undergoes a rapid change of direction
around $\tau=0$, assuming $\epsilon$ to be small. The form of
$f(\tau)$ acts as a regulator needed to avoid pathologies in the
calculation that follows. However, dependence on this regulator
cancels out, so that an explicit form for $f(\tau)$ will not be
needed. Owing to the constant nature of $u$ and $u'$, the acceleration
is given by
\begin{equation}
\dot{\lambda}^\mu=\dot{f}(\tau)\left({u'}^{\mu}-u^\mu\right).
\label{lambdadot}
\end{equation}
The acceleration vanishes for $\tau<-\epsilon$ and $\tau>\epsilon$, but is
potentially large in the interval $(-\epsilon,\epsilon)$. Without loss
of generality, we may choose the spatial origin to be the place at
which the particle changes direction, so that $y^\mu(0)=0$.

\subsection{Gauge theory}

We first consider the gauge theory case, and start by using the
definitions of eqs.~(\ref{kdef}) to write the current density of
eq.~(\ref{jadef}) as
\begin{equation}
j^{\nu}_\textrm{\tiny{KS}}=\frac{2g}{4\pi}\int d\tau\frac{\dot{\lambda}(\tau)
\cdot(x-y(\tau))}{[\lambda(\tau)\cdot (x-y(\tau))]^4}(x-y(\tau))^\nu
\delta(\tau-\tau_{\rm ret}),
\label{janu1}
\end{equation}
where we have introduced a delta function to impose the retarded time
constraint. Using the identity
\begin{equation}
\frac{\delta(\tau-\tau_{\rm ret})}{\lambda\cdot (x-y(\tau))}=
2\theta(x^0-y^0(\tau))\delta\left((x-y(\tau))^2\right),
\label{deltaid}
\end{equation}
one may rewrite eq.~(\ref{janu1}) as
\begin{equation}
j^{\nu}_\textrm{\tiny{KS}}=\frac{4g}{4\pi}\int d\tau\frac{\dot{\lambda}(\tau)
\cdot(x-y(\tau))}{[\lambda(\tau)\cdot (x-y(\tau))]^3}(x-y(\tau))^\nu
\theta(x^0-y^0(\tau))\delta\left((x-y(\tau))^2\right).
\label{janu2}
\end{equation}
Any radiation field will be associated with the non-zero
acceleration only for $|\tau|<\epsilon$, where $y^\mu(\tau)$ is small. We
may thus neglect this with respect to $x^\mu$ in
eq.~(\ref{janu2}). Substituting eq.~(\ref{lambdadot}) then gives
\begin{equation}
j^{\nu}_\textrm{\tiny{KS}}=\frac{4g}{4\pi}x^\nu\theta(x^0)\delta(x^2)
\int_{-\epsilon}^\epsilon d\tau\frac{b \dot{f}(\tau)}{(a+bf(\tau))^3},
\label{janu3}
\end{equation}
where
\begin{equation}
a=x\cdot u,\quad b=x\cdot u'-x\cdot u.
\label{abdef}
\end{equation}
The integral is straightforwardly carried out to give 
\begin{align}
j^{\nu}_\textrm{\tiny{KS}}&=-\frac{2g}{4\pi}x^\nu\theta(x^0)\delta(x^2)
\left[\frac{1}{(x\cdot u')^2}-\frac{1}{(x\cdot u)^2}\right]\notag\\
&=\frac{2g}{4\pi}\theta(x^0)\delta(x^2)
\left[\frac{\partial}{\partial u'_\nu}\left(\frac{1}{x\cdot u'}\right)
-(u'\rightarrow u)\right].
\label{janu4}
\end{align}
One may now Fourier transform this expression, obtaining a current depending on a momentum $k$ conjugate to the position $x$. As our aim is to extract a scattering amplitude from the Fourier space current, $\tilde j^\mu_\textrm{\tiny{KS}}(k)$, we consider only the on-shell limit of the current where $k^2 = 0$; we also drop terms in $\tilde j^\mu_\textrm{\tiny{KS}}(k)$ which are proportional to $k^\mu$ as these terms are pure gauge. The technical details are presented in appendix~\ref{app:Fourier}, and the result is
\begin{equation}
\tilde{j}^{\nu}_\textrm{\tiny{KS}}(k)=-ig\left(\frac{{u'}^\nu}
{u'\cdot k}-\frac{u^\nu}{u\cdot k}\right).
\label{jafourier}
\end{equation}
We may now interpret this as follows. First, we note that the current
results upon acting on the radiative gauge field with an inverse
propagator, consistent with the LSZ procedure for truncating Green's
functions. It follows that the contraction of $\tilde{j}^{\nu}_\textrm{\tiny{KS}}$ with a
polarisation vector gives the scattering amplitude for emission of a
gluon. Upon doing this, one obtains the standard eikonal scattering amplitude for
Bremsstrahlung (see e.g.~\cite{Peskin:1995ev})
\begin{equation}
{\cal A}_{\rm gauge}\equiv  \epsilon_\nu(k)\tilde{j}^{\nu}_\textrm{\tiny{KS}}
=-i g\left(\frac{\epsilon\cdot u'}
{u'\cdot k}-\frac{\epsilon\cdot u}{u\cdot k}\right) \,.
\label{Agauge}
\end{equation}
We thus see directly
that the additional current density in the Kerr-Schild approach
corresponds to the radiative part of the gauge field.

\subsection{Gravity}

We now turn to the gravitational case. Our goal is to extract the eikonal scattering amplitude for gravitational Bremsstrahlung from the Kerr-Schild stress-energy tensor $T_\textrm{\tiny{KS}}^{\mu\nu}$ for a particle of mass $M$ moving along precisely the same trajectory as our point charge. Thus, the acceleration of the particle is, again,
\begin{equation}
\dot{\lambda}^\mu=\dot{f}(\tau)\left({u'}^{\mu}-u^\mu\right).
\end{equation}
The calculation is a precise parallel to the calculation of the Bremsstrahlung amplitude for the point charge. However, as we will see, the presence of an additional factor of the Kerr-Schild vector $k^\nu$ in the gravitational case leads to a slightly different integral which we encounter during the calculation. This integral cancels the factor of 3 which appears in $T_\textrm{\tiny{KS}}^{\mu\nu}$, restoring the expected numerical factors in the momentum space current. Let us now turn to the explicit calculation. \\

We begin by writing the stress tensor as an integral over a delta function which enforces the retardation and causality constraints
\begin{equation}
T^{\mu\nu}_\textrm{\tiny{KS}}=\frac{3M}{2\pi}\int d\tau\frac{\dot{\lambda}(\tau)
\cdot(x-y(\tau))}{[\lambda(\tau)\cdot (x-y(\tau))]^4}(x-y(\tau))^\mu(x-y(\tau))^\nu
\theta(x^0-y^0(\tau))\delta\left((x-y(\tau))^2\right),
\end{equation}
corresponding to eq.~\eqref{janu2} in the gauge theoretic case. The fourth power in the denominator in the gravitational case arises as a consequence of the additional factor of $k^\mu = (x-y(\tau))^\mu / [\lambda(\tau)\cdot (x-y(\tau))]$. As before, the integral is strongly peaked around $y^\mu = 0$, and we may perform the integral in this region to find that
\begin{align}
T^{\mu\nu}_\textrm{\tiny{KS}}&=-\frac{2M}{4\pi}x^\mu x^\nu\theta(x^0)\delta(x^2)
\left[\frac{1}{(x\cdot u')^3}-\frac{1}{(x\cdot u)^3}\right]\notag\\
&=-\frac{M}{4\pi}\theta(x^0)\delta(x^2)\left[\frac{\partial}
{\partial u'_\mu}
\frac{\partial}{\partial u'_\nu}\left(\frac{1}{x\cdot u'}\right)
-(u'\rightarrow u)\right].
\label{Tmunurad2}
\end{align}

Notice that the factor 3 in the numerator of the stress-energy tensor has cancelled due to the additional factor of $\lambda(\tau)\cdot (x-y(\tau))$ in the denominator of the integrand in the gravitational case. The double copy structure is evidently now captured by a replacement of one derivative $\frac{\partial}{\partial u'_\nu}$ in gauge theory with two derivatives $\frac{\partial}{\partial u'_\mu} \frac{\partial}{\partial u'_\nu}$ in gravity. \\

Our next step is to Fourier transform to momentum space. The calculation is extremely similar to the gauge theoretic case (again, see
appendix~\ref{app:Fourier}). As our goal is to compute a scattering amplitude, we work in the on-shell limit $k^2 = 0$ and omit pure gauge terms. After a short calculation, we find
\begin{equation}
\tilde{T}^{\mu\nu}_\textrm{\tiny{KS}}(k)=-iM\left(\frac{{u'}^\mu{u'}^\nu}
{u'\cdot k}-\frac{u^\mu u^\nu}{u\cdot k}\right).
\label{Tmunufourier}
\end{equation}
To construct the scattering amplitude, we must 
contract this Fourier-transformed stress-energy tensor with a polarisation tensor, which may be written as an outer product of two gauge theory polarisation vectors:
\begin{equation}
\epsilon^{\mu\nu}(k)=\epsilon^\mu(k)\epsilon^\nu(k).
\label{epstens}
\end{equation}
The scattering amplitude is then given by 
\begin{equation}
{\cal A}_{\rm grav}\equiv \epsilon_\mu(k)\epsilon_\nu(k)
\tilde{T}^{\mu\nu}_\textrm{\tiny{KS}}(k)= -i M\left(\frac{\epsilon\cdot u'
\,\epsilon\cdot {u'}}
      {u'\cdot k}-\frac{\epsilon\cdot u\, \epsilon\cdot u}{u\cdot k}\right),
\label{Agrav}
\end{equation}
corresponding to the known eikonal amplitude for gravitational
Bremsstrahlung~\cite{Weinberg:1965nx}. Again we see that the additional
source term in the Kerr-Schild approach corresponds to the radiative
part of the field. Furthermore, in this form the standard double
copy for scattering amplitudes is manifest: numerical factors agree between
eqs.~(\ref{jafourier}) and~(\ref{Tmunufourier}), such that the mass in
the gravity theory is replaced with the colour charge in the gauge
theory, as expected from the usual operation of the classical single
copy~\cite{Monteiro:2014cda,Luna:2015paa}. \\

Let us summarise the results of this section. We have examined the
particular case of a particle which undergoes a rapid change in
direction, and confirmed that the additional source terms appearing in
the Kerr-Schild description (in both gauge and gravity theory) are
exactly given by known radiative scattering amplitudes. This directly links the classical double copy to the BCJ procedure for
amplitudes. \\

It is interesting to compare the BCJ double copy for scattering amplitudes with the Kerr-Schild double copy, which has been formulated in position space. It is clear that momentum space is the natural home of the double copy. For scattering amplitudes, the amplitudes themselves and the double copy procedure are local in momentum space. In our Bremsstrahlung calculation, the numerical coefficients in the sources are also more natural after the Fourier transform. On the other hand, the currents $T^{\mu\nu}_\textrm{\tiny{KS}}$ and $j^{\nu}_\textrm{\tiny{KS}}$ are also local in position space. This unusual situation arises because the scattering amplitudes do not conserve momentum: in any Bremsstrahlung process, some momentum must be injected in order to bend the point particle trajectory. Of course, in the case of a static point particle locality in both position space and momentum space is more natural. This is reflected by the structure of the Fourier transform in the present case: as explained in Appendix~\ref{app:Fourier}, the factor $1/x\cdot u$ describing a particle worldline Fourier transforms to an integrated delta function $\int_0^\infty dm \; \delta^4(q - m u)$ (see eq.~\eqref{jfourier3}).

\section{Gravitational energy conditions}
\label{sec:energy}

In this section, we consider the null, weak and strong energy conditions
of general relativity. These were recently examined in the context of
the Kerr-Schild double copy in ref.~\cite{Ridgway:2015fdl}, where it
was shown that extended charge distributions double copy to matter
distributions that cannot simultaneously obey the weak and strong energy
conditions, if there are no spacetime singularities or horizons. Although the point
particle solution of interest to us has both singularities and horizons, it is
still interesting to examine the energy conditions. \\

\begin{figure}
\begin{center}
\scalebox{0.7}{\includegraphics{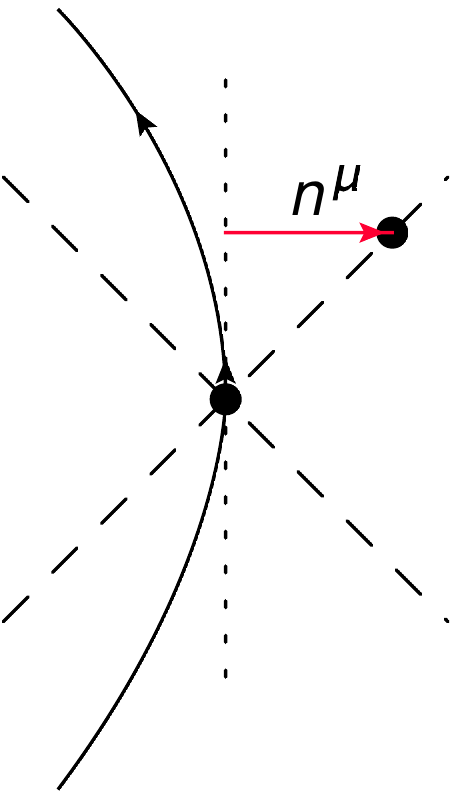}}
\caption{Physical interpretation of $(k\cdot \dot{\lambda})$, where this
  denotes the component of acceleration in the direction $n^\mu$.}
\label{fig:worldline2}
\end{center}
\end{figure}

The null energy
condition on a given energy-momentum tensor can be expressed by
\begin{equation}
T_{\mu\nu}\ell^\mu \ell^\nu\geq 0,
\label{nullenergy}
\end{equation}
where $\ell^\mu$ is any future-pointing null vector.
The weak energy
condition is similarly given by
\begin{equation}
T_{\mu\nu}t^\mu t^\nu\geq 0,
\label{weakenergy}
\end{equation}
for any future-pointing timelike vector $t^\mu$. The interpretation of
this condition is that observers see a non-negative matter density. The null energy condition is implied by the weak energy condition (despite the names, the former is the weakest condition).
One may also stipulate that the trace of the tidal tensor
measured by such an observer is non-negative, which leads to the
strong energy condition
\begin{equation}
T_{\mu\nu}t^\mu t^\nu\geq \frac{T}{2}g_{\mu\nu} t^\mu t^\nu,\quad
T\equiv T^\alpha_\alpha .
\label{strongenergy}
\end{equation}
Let us now examine whether these conditions are satisfied by the
Kerr-Schild energy-momentum tensor of eq.~(\ref{Tmunurad}). First, the
null property of the vector $k^\mu$ implies that the trace vanishes,
so that the weak and strong energy conditions are equivalent. We may further unify these with the null energy condition, by noting that eq.~(\ref{Tmunurad}) implies
\begin{equation}
T^{\mu\nu}_\textrm{\tiny{KS}}V_\mu V_\nu=(k\cdot\dot{\lambda})
\left[\frac{3M (k\cdot V)^2}{4\pi r^2}\right].
\label{KSenergy}
\end{equation}
for {\it any} vector $V^\mu$. The quantity in the square brackets is
positive definite, so that whether or not the energy conditions are
satisfied is purely determined by the sign of $k\cdot \dot{\lambda}$.
This scalar quantity is easily determined in the instantaneous rest-frame
of the point particle; it is the negative of the component of acceleration
in the direction $n^\mu$ of the observer (at the retarded time), see figure~\ref{fig:worldline2}.
Thus the energy conditions are not satisfied throughout the spacetime. In particular, any observer which sees the particle accelerating towards (away from) her will measure a negative (positive) energy density. \\

We remind the reader that the energy-momentum tensor is, in the case under study, an effective way of representing the full vacuum solution. The latter will have no issues with energy conditions. Analogously, the Li\'enard-Wiechert vacuum solution in gauge theory can be represented, as we have shown in Section~\ref{sec:KSradiate}, by a boosted Coulomb field, together with a charged current encoding the radiation.

\section{Discussion}
\label{sec:conclude}

In this paper, we have extended the classical double copy of
refs.~\cite{Monteiro:2014cda,Luna:2015paa} to consider
accelerating, radiating point sources. This significantly develops previous
results, which were based on stationary Kerr-Schild solutions, to a situation involving explicit time dependence. 
The structure of the double copy we have observed in the radiating case is
precisely as one would expect. Passing from the gauge to the gravity theory,
the overall scalar function $\phi$ is left intact; indeed it is the well-known scalar
propagator in four dimensions. This is the same as the treatment of scalar propagators
in the original BCJ double copy procedure for amplitudes. Similarly, the tensor structure 
of the gravitational field is obtained from the gauge field by replacing the vector $k_\mu$ by
the symmetric, trace-free tensor $k_\mu k_\nu$. Finally, our use of Kerr-Schild coordinates
in gravity linearised the Einstein tensor (with mixed indices). Reflecting this linearity,
the associated single copy satisfies the linearised Yang-Mills equations. \\

It is worth dwelling a little on the physical implication of our work. The classical double copy is known
to relate point sources in gauge theory to point sources in general relativity, in accordance
with intuition arising from scattering amplitudes. In this article, we have simply considered the case
where the point sources move on a specified, arbitrarily accelerated, timelike worldline. On general
grounds we expect radiation to be emitted due to the acceleration. Our use of Kerr-Schild coordinates
organised the radiation into sources appearing on the right-hand side of the field equations: a current
density in gauge theory, and a stress-energy tensor in gravity. Intriguingly, we found that the expressions for these
sources also have a double copy structure: one passes from the gauge current to the gravitational stress-energy
tensor by replacing $k_\mu$ by $k_\mu k_\nu$ while leaving a scalar factor intact, up to numerical factors
which are canonical in momentum space. Since these sources encode the complete radiation fields
for the accelerating charge and black hole, there is a double copy between the radiation generated by
these two systems. This double copy is a property of the exact solution of gauge theory and general relativity. \\

We further extracted one simple perturbative scattering amplitude from this radiation field, namely the Bremsstrahlung 
scattering amplitude. The double copy property was maintained as we extracted the scattering amplitude, which firmly establishes a link between the double copy for amplitudes and the double copy for classical solutions. \\

However, we should emphasise one unphysical aspect of our setup. We mandated a wordline for our point particle
in both gauge theory and general relativity. In gauge theory, this is fine: one can imagine that an external force
acts on the particle causing its worldline to bend. However, in general relativity such an external force would contribute
to the stress-energy tensor in the spacetime. Since we ignored this component of the stress-energy tensor, our calculation
is not completely physical. Instead, one should regard the point particle in both cases as a specified boundary condition,
rather than as a physical particle. We have therefore seen that the radiation generated by this boundary condition enjoys
a precise double copy. \\

There are a number of possible extensions of our results. One may look
at time-dependent extended sources in the Kerr-Schild description, for
example, or particles accelerating in non-Minkowski backgrounds (for
preliminary work in the stationary case, see
ref.~\cite{Luna:2015paa}). It would also be interesting to examine
whether a double copy procedure can be set up in other coordinate
systems, such as the more conventional de Donder gauge. 
One particularly important issue is to understand the generalisation of the colour-dual
requirement on kinematic numerators to classical field backgrounds. The Jacobi relations satisfied by colour-dual numerators
hint at the existence of a kinematic algebra~\cite{Monteiro:2011pc,Fu:2016plh} underlying the connection between gauge theory
and gravity; revealing the full detail of this structure would clearly be an important breakthrough. The study of
the classical double copy is in its infancy, and many interesting avenues have yet to be explored.

\section*{Acknowledgments}

We thank John Joseph Carrasco and Radu Roiban for many illuminating discussions and thought-provoking questions. CDW is supported by the UK Science and Technology Facilities Council
(STFC) under grant ST/L000446/1, and is perennially grateful to the
Higgs Centre for Theoretical Physics for hospitality. DOC is supported
in part by the STFC consolidated grant ``Particle Physics at the Higgs
Centre'', by the National Science Foundation under grant NSF
PHY11-25915, and by the Marie Curie FP7 grant 631370. AL is supported
by Conacyt and SEP-DGRI studentships.

\appendix

\section{Fourier transform of source terms}
\label{app:Fourier}

In this appendix, we describe how to carry out the Fourier transform
of eqs.~(\ref{janu4}, \ref{Tmunurad2}), to get the momentum-space
expressions of eqs.~(\ref{jafourier}, \ref{Tmunufourier}).\\

One may first consider the transform of $(u\cdot x)^{-1}$, where we
work explicitly in four spacetime dimensions:
\begin{align}
{\cal F}\left\{\frac{1}{u\cdot x}\right\}&=\int d^4 x\frac{e^{iq\cdot x}}
{u\cdot x}\notag\\
&=\frac{1}{u^0}\int d^3 x e^{-i\vec{q}\cdot \vec{x}}\int dx^0\frac{e^{iq^0x^0}}
{x^0-\frac{\vec{x}\cdot \vec{u}}{u^0}}.
\label{jfourier1}
\end{align}
Closing the $x^0$ contour in the upper half plane gives a
positive frequency solution $q^0>0$:
\begin{align}
{\cal F}\left\{\frac{1}{u\cdot x}\right\}&=\frac{2\pi i}{u^0}
\int d^3x \, e^{-i\vec{x}\cdot\left[\vec{q}-\frac{q^0}{u^0}\vec{u}\right]}\notag\\
&=\frac{i(2\pi)^4}{u^0}\delta^{(3)}\left(\vec{q}-\frac{q^0}{u^0}\vec{u}\right).
\label{jfourier2}
\end{align}
It is possible to regain a covariant form for this expression by introducing a
mass variable $m$, such that
\begin{align}
{\cal F}\left\{\frac{1}{u\cdot x}\right\}&=\frac{i(2\pi)^4}{u^0}
\int_0^\infty dm\, \delta\left(m-\frac{q^0}{u^0}\right)\delta^{(3)}(\vec{q}
-m\vec{u})\notag\\
&=i(2\pi)^4\int_0^\infty dm \, \delta^{(4)}(q-mu),
\label{jfourier3}
\end{align}
where the integral is over non-negative values of $m$ only, given that
$q^0>0$. Given that $\theta(x^0)\delta(x^2)$ is a retarded
propagator\footnote{The retarded nature of the propagator is
  implemented by the prescription
  $\frac{1}{(p^0+i\varepsilon)^2-\vec{p}^2}$, where $\varepsilon$
  ensures convergence of the integrals in what follows.}, one may also
note the transform 
\begin{equation}
{\cal F}\left\{\theta(x^0)\delta(x^2)\right\}=-\frac{2\pi}{q^2}.
\label{proptrans}
\end{equation}
We then use the convolution theorem to obtain the Fourier transform of the current from eq. (\ref{janu4}). The theorem states that the Fourier transform of a product is equal to the convolution of the transforms of each term. That is, 
\begin{eqnarray}
\mathcal{F}\{f\cdot g\}=\mathcal{F}\{f\}\ast\mathcal{F}\{g\},
\end{eqnarray} 
where the convolution operation in four dimensions takes the form
\begin{eqnarray}
(F\ast G)(k)=\frac{1}{(2\pi)^4}\int d^4qF(q)G(k-q).
\label{conv}
\end{eqnarray}
Then, we can compute the Fourier transform of the current
\begin{align}
\nonumber\tilde{j}^{\nu}(k)&=\mathcal{F}\{j_\textrm{\tiny{KS}}^\nu(x)\}\\
&=\frac{2g}{4\pi}
\frac{\partial}{\partial u'_\nu}\left[\mathcal{F}\{\theta(x^0)\delta(x^2)\}\ast\mathcal{F}\left\lbrace\frac{1}{x\cdot u'}\right\rbrace\right]
-(u\leftrightarrow u'),
\end{align}
so inserting eqs. (\ref{proptrans}) and (\ref{jfourier3}), and using the convolution definition eq. (\ref{conv}) we obtain the expression
\begin{align}
\nonumber\tilde{j}^{\nu}(k)&=\frac{2g}{4\pi}
\frac{\partial}{\partial u'_\nu}\left[\frac{1}{(2\pi)^4}\int d^4q\left(-\frac{2\pi}{q^2}\right)\left(i(2\pi)^4\int_0^\infty dm \, \delta^{(4)}(k-q-mu')\right)\right]
-(u\leftrightarrow u')\\
&=-ig\int_0^\infty dm\left(\frac{\partial}{\partial u'_\nu}
\left[\frac{1}{(k-mu')^2}\right]-(u\leftrightarrow u')\right).
\label{jfourier4}
\end{align}
where we have carried out the integral over $q$ in the
last line. The derivative in the $m$ integral can be carried out to
give
\begin{equation}
\int_0^\infty dm \frac{2m(k-mu')^\nu}{(k-mu')^4}=-\int_0^\infty dm
\frac{2m^2 {u'}^\nu}{(m^2-2mu'\cdot k)^2},
\label{mint1}
\end{equation}
where, on the right-hand side, we have used the onshellness condition $k^2=0$, and also
neglected terms $\sim k^\mu$, which vanish upon contraction of the current with a
physical polarisation vector. The remaining integral
over $m$ is easily carried out, and leads directly to the result of
eq.~(\ref{jafourier}). \\

Similar steps to those leading to eq.~(\ref{jfourier4}) can be used to
rewrite eq.~(\ref{Tmunurad2}) in the form
\begin{equation}
T^{\mu\nu}_\textrm{\tiny{KS}}=\frac{iM}{2} \int_0^\infty dm\left(\frac{\partial}{\partial
u'_\mu}\frac{\partial}{\partial u'_\nu}\left[\frac{1}{(k-mu')^2}\right]
-(u\leftrightarrow u')\right).
\label{Tmunufourier1}
\end{equation}
Carrying out the double derivative gives
\begin{align}
\frac{\partial}{\partial
u'_\mu}\frac{\partial}{\partial u'_\nu}\left[\frac{1}{(k-mu')^2}\right]
&=-\frac{2m^2\eta^{\mu\nu}}{(m^2-2mu'\cdot k)^4}+\frac{8m^2(k-mu')^\mu
(k-mu')^\nu}{(m^2-2mu'\cdot k)^3}\notag\\
&\simeq \frac{8m^4{u'}^\mu{u'}^\nu}{(m^2-2mu'\cdot k)^3},
\label{Tmunufourier2}
\end{align}
where in the second line we have again used onshellness ($k^2=0$), and
ignored terms which vanish when contracted with the graviton
polarisation tensor. Substituting eq.~(\ref{Tmunufourier2}) into
eq.~(\ref{Tmunufourier1}), the $m$ integral is straightforward, and
one obtains the result of eq.~(\ref{Tmunufourier}).

\bibliography{refs.bib}
\end{document}